\PassOptionsToPackage{hyphens}{url}
\documentclass{article}

\usepackage{PRIMEarxiv}

\usepackage[utf8]{inputenc} 
\usepackage[T1]{fontenc}    
\usepackage{hyperref}       
\usepackage{url}            
\usepackage{booktabs}       
\usepackage{tabularx}       
\usepackage{amsmath,amssymb,amsfonts}
\usepackage{nicefrac}       
\usepackage{microtype}      
\usepackage{fancyhdr}       
\usepackage{graphicx}       
\graphicspath{{media/}{./}} 

\title{AI for eGovernance: Leveraging Language Technologies for Digital Inclusion
}

\author{
  Cynthia Jayne Amol \\
  Maseno University \\
  Maseno, Kenya \\
  \texttt{cynthia@maseno.ac.ke} \\
  \And
  Lilian Wanzare \\
  Maseno University \\
  Maseno, Kenya \\
  \texttt{ldwanzare@maseno.ac.ke} \\
}

\begin{document}
\maketitle

\begin{abstract}
In a bid to champion Africa's digital transformation as outlined in the Kenya Digital Master Plan 2022-2032, the Kenyan government has invested heavily in the digitisation of public service delivery. The national eGovernment platform, eCitizen, hosts an array of services that are vital to citizens' livelihoods, including business registration, youth enterprise lending, digital literacy, and tax filing. However, the platform faces digital inclusion challenges that often limit equitable access for marginalised groups, particularly youth in rural areas and persons with disabilities. This exclusion is driven by a language barrier, since the platform is primarily in English with limited Swahili support, and by a text-only interface that excludes low-literacy and visually impaired users. This paper reports findings from six one-on-one interviews and an eleven-person roundtable discussion across four platforms considered essential to citizens' livelihoods, with a particular focus on youth: Ajira Digital, Kenya Revenue Authority (KRA) iTax, Higher Education Loans Board (HELB), and the Youth Enterprise Development Fund (YEDF). Drawing on these findings, on Web Content Accessibility Guidelines (WCAG) 2.1 principles as a design lens for diagnosing eCitizen's accessibility gaps, and on comparator models such as Estonia's \textit{B\"urokratt} and India's \textit{Bhashini}, the paper proposes a human-centered, systems-minded framework for embedding Generative AI, language technologies, including text simplification, neural machine translation, text-to-speech and speech-to-text, into eCitizen. The paper argues that, when deployed responsibly and supported by appropriate human oversight, these interventions could meaningfully improve access to government services for linguistically excluded citizens. However, it also emphasises that AI alone cannot overcome the broader structural barriers, including unemployment and educational inequality, that also shape digital exclusion.
\end{abstract}

\keywords{eGovernance \and digital inclusion \and language technologies \and accessibility \and generative AI \and machine translation \and Kenya}

\section{Introduction}
\label{sec:introduction}
Efficient public service delivery is an essential part of good governance. In Africa, developing countries like Kenya are constantly leveraging emerging technology to automate services in order to compete globally. The Kenya Digital Master Plan 2022--2032 \cite{icta2022} documents this investment in eGovernance, aimed at improving service delivery efficiency and streamlining information access, an effort that affects citizens' livelihoods. This builds on earlier policy foundations such as Kenya's first eGovernment Strategy \cite{govkenya2004}, which established the administrative structure, training requirements, and standardisation framework that later digitisation efforts, including eCitizen, have built upon.

eGovernance uses Information Communication and Technologies (ICTs), such as online portals, to deliver public services \cite{malodia2021}, replacing the manual systems that previously caused delays and a lack of transparency. Digital service delivery through eGovernment platforms ought to be fast, user-friendly and inclusive to all citizens \cite{srinivas2025}, fulfilling the government's mandate of equitable governance. In Kenya, the central eGovernment platform, eCitizen\footnote{\url{https://accounts.ecitizen.go.ke/en}}, hosts over 22,000 essential government services, including business and company registrations, tax services, Higher Education Loans Board (HELB) applications, Ajira Digital for digital education, passport or visa applications and driver's licence renewal. These services are vital to all citizens' livelihoods, especially the youth, who make up over 80\% of the population and are key drivers of the economy \cite{ncpd2021}.

\begin{figure}[htbp]
  \centering
  \includegraphics[width=0.8\linewidth]{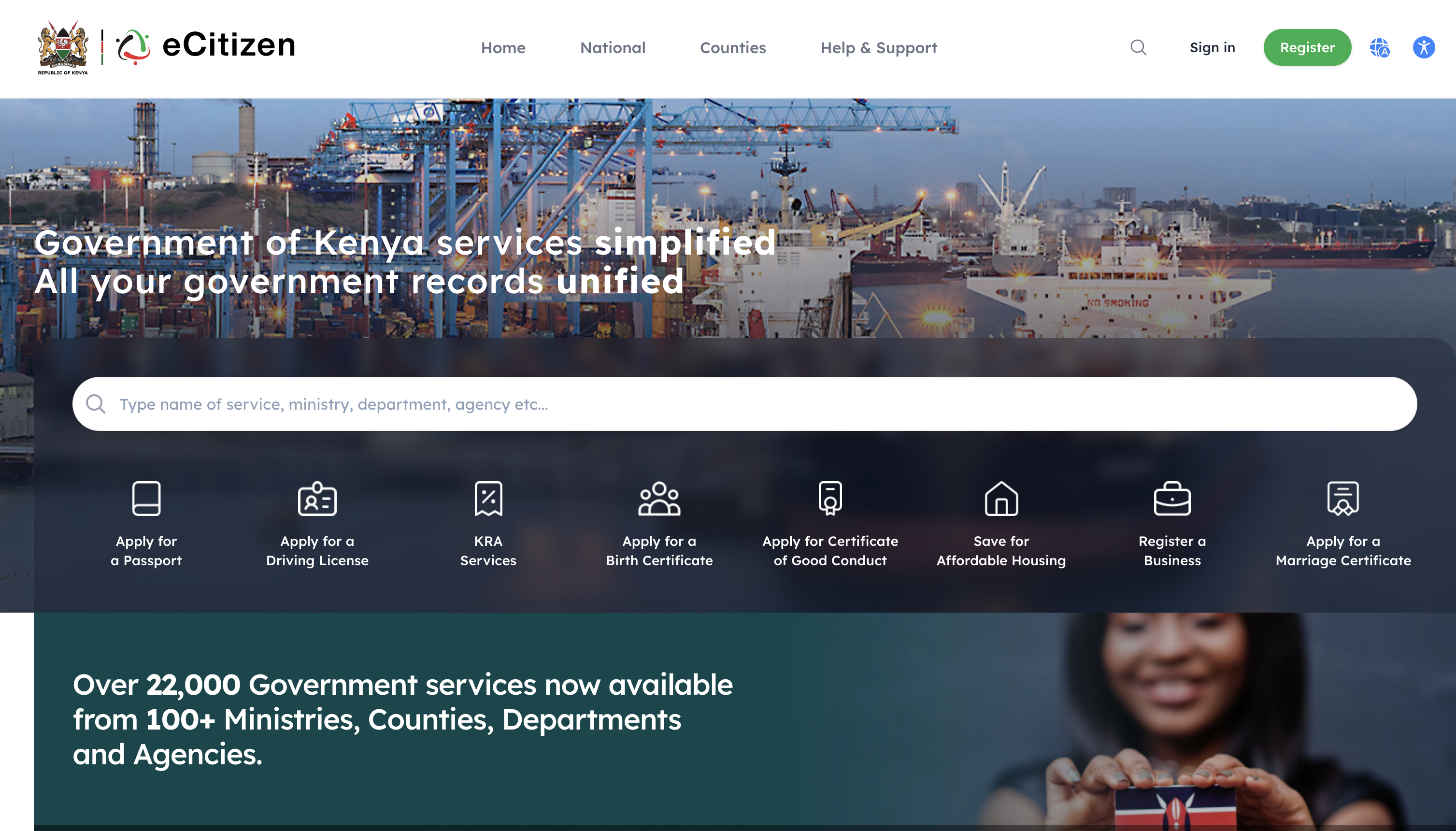}
  \caption{Default (English) eCitizen landing page.}
  \label{fig:homepage_en}
\end{figure}

\begin{figure}[htbp]
  \centering
  \includegraphics[width=0.8\linewidth]{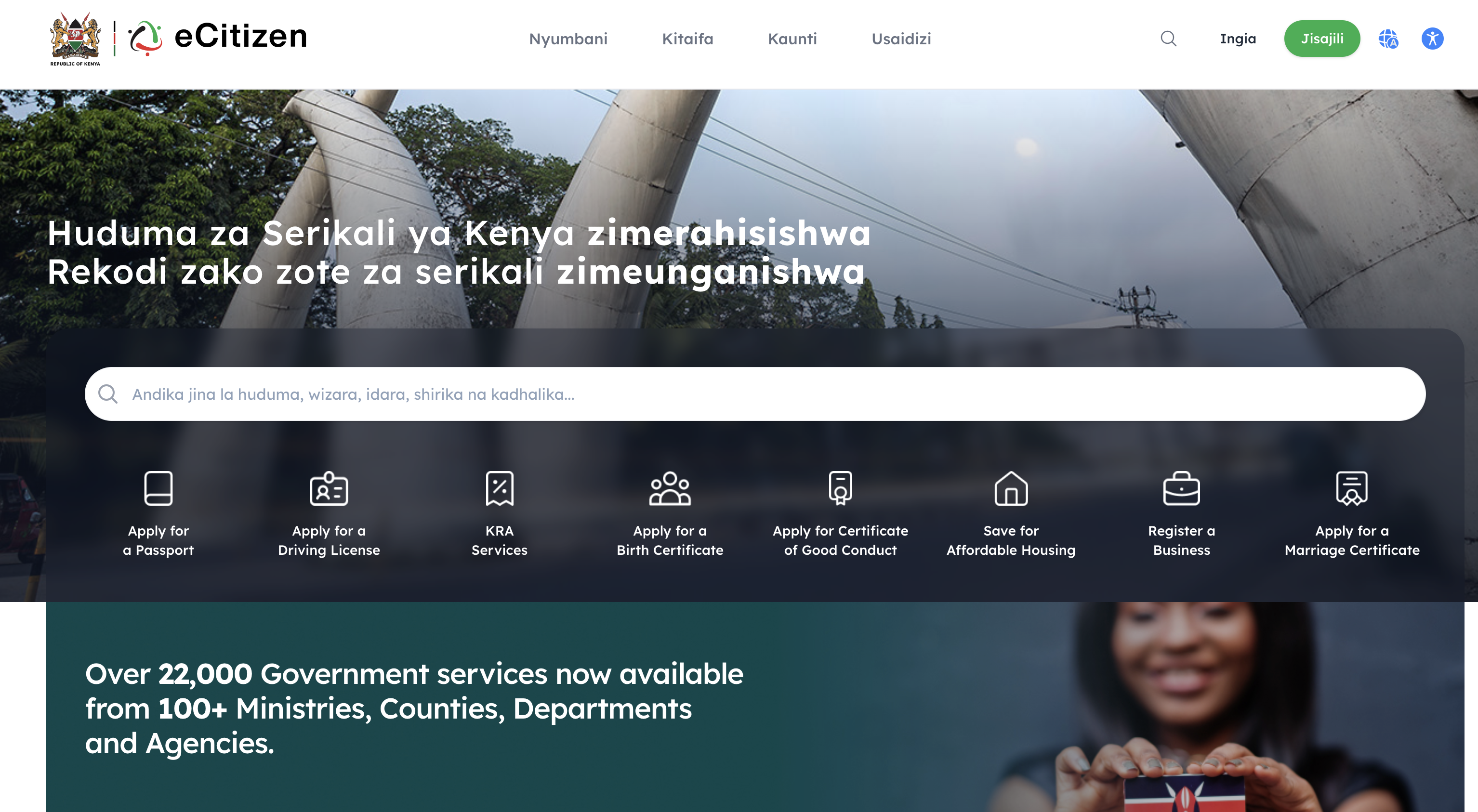}
  \caption{Translated (Swahili) eCitizen landing page.}
  \label{fig:homepage_sw}
\end{figure}

However, eCitizen has not yet achieved absolute digital inclusion. It limits equitable access to services, particularly for users with limited mastery of national languages and persons with disabilities \cite{djatmiko2025}. A key barrier lies in the platform's linguistic limitations. It is primarily in English, with Swahili support typically limited to webpage headers, as shown in Figure~\ref{fig:homepage_en} and Figure~\ref{fig:homepage_sw}. The main content stays in English despite Kenya's linguistic diversity of over sixty native and eight non-native languages \cite{barasa2023}. This language gap restricts access to services that could unlock education, employment, entrepreneurship and social welfare, all of which directly improve livelihoods. The lack of audio support further alienates citizens who cannot read.

In Kenya, English is often learned as a third language, after an individual's native ethnic language and Swahili \cite{kerkhoff2022}. English serves as the official language for formal government and legal contexts, but Swahili functions as a cross-cultural lingua franca widely spoken among youth in both urban and rural settings \cite{luvaga2021}, and in many rural areas, native languages are the everyday language, not English or Swahili \cite{barasa2023,kerkhoff2022}. This layered linguistic landscape means the eCitizen platform's design alienates a significant portion of the population, particularly youth and those in rural communities. The government must move beyond translating headers and meaningfully integrate Swahili and other native languages, letting all citizens, regardless of literacy or physical ability, access and benefit from eGovernance services, and fulfilling the platform's potential as a tool for social and economic empowerment.

Artificial Intelligence (AI)-driven language technologies offer an interim way to close this gap. As AI for low-resource languages advances, tools such as neural machine translation (NMT), text summarisation, and chatbot-based user support can expand access to digital government services. Recent research \cite{gitau2023,wanjawa2022} has produced language resources for Swahili and other Kenyan languages, making such solutions increasingly feasible to implement.

This paper first situates eCitizen among comparable empirical studies and existing eGovernment platforms, such as Estonia's \textit{B\"urokratt} and India's \textit{Bhashini}. It then synthesises interviews and a roundtable held with users and support staff of Ajira Digital, KRA iTax, HELB and YEDF. These findings are discussed alongside Kenya's and the African Union's AI strategies and a WCAG-based review of eCitizen's design. The findings inform the technical and governance recommendations proposed for embedding language technologies into eCitizen to improve digital inclusion.

\section{Related Work}
\label{sec:related_work}
This section reviews prior research on language and accessibility barriers in eGovernment, existing eGovernment technology models, and the language technology resources available for Kenyan languages.

\subsection{Language and Accessibility Barriers in eGovernment}
Language exclusion in eGovernment is a recurring challenge across postcolonial Africa. Many African states inherited their official languages from colonial administrations, and digital government platforms have largely continued to operate within these frameworks, with limited support for indigenous languages. In South Africa, indigenous-language users face mistranslated policy terms and rely on literate relatives or intermediaries to complete online processes \cite{ramothwa2026}. These same patterns, task delegation and uncertainty from unclear terminology, appear in the discussions from interviews presented in Section~\ref{sec:findings}. Language is only one factor shaping access to eGovernment services. Adoption is also influenced by social, infrastructural and trust-related factors alongside interface and usability challenges \cite{djatmiko2025}. Among older citizens in Kenya, non-use has been linked to digital literacy, self-efficacy and infrastructure limitations, rather than platform design alone \cite{muchiri2025}. Physical alternatives provide only partial solutions. Huduma Centres, set up to support citizens who face barriers to digital access, are unevenly distributed, particularly in rural and peri-urban areas \cite{nkanata2022}. Similar access and participation challenges have been reported in Ugandan local government \cite{michael2023}, suggesting language exclusion is part of a wider regional pattern in the design of inclusive digital public services.

\subsection{Comparator eGovernment Technology Models}
The approach proposed in this paper draws on established eGovernment models from Africa and beyond. Rwanda's IremboGov\footnote{\url{https://irembo.gov.rw/}}, a centralised digital service platform, combines online services with an agent network to support citizens who face digital literacy barriers, and its localisation into Kinyarwanda shows the value of designing for users' linguistic contexts. South Africa offers a policy example through its framework of functional multilingualism. With eleven official languages, government institutions are expected to incorporate local languages into public communication and service delivery. Together, these examples show that inclusive digital government needs both accessible platforms and the institutional frameworks to support them.

Globally, AI-enabled government systems are exploring conversational interfaces and automated assistance. Estonia's B\"urokratt\footnote{\url{https://www.kratid.ee/burokratt}} is an AI virtual assistant that connects government databases so users can file complaints or renew identification documents in one place \cite{oecdai2025}. India's Bhashini\footnote{\url{https://bhashini.gov.in/}} initiative develops language technologies for translation and multimodal interaction across Indian languages \cite{meity2025}. Both offer a useful reference for digital public services that combine AI assistance with multilingual interaction.

\subsection{Language Technology Landscape}
Multilingual eGovernment services need language resources such as parallel corpora, annotated speech datasets, and pretrained translation and speech models. Large language models (LLMs) can support clearer communication by adapting complex information to different literacy needs. Johnson et al.\ \cite{johnson2025} found that AI-simplified textbook content improved reading efficiency and comprehension for participants with reading challenges. The study concerns educational rather than government text, but it is a useful reference point for simplifying public information, including business registration, tax filing and HELB applications.

Kenya's language technology ecosystem is still at an earlier stage than larger national initiatives. Masakhane Web \cite{gitau2023} provides open machine translation resources for African languages including Swahili, while Kencorpus \cite{wanjawa2022} provides annotated corpora for Swahili, Dholuo and Luhya that could support future translation and speech technologies. Compared with Bhashini's national-scale language infrastructure or B\"urokratt's integrated architecture, Kenya's language resources are limited, with implications for how quickly reliable multilingual support could be deployed on platforms such as eCitizen.

\section{Methodology}
\label{sec:methodology}
\begin{figure}[htbp]
  \centering
  \includegraphics[width=0.8\linewidth]{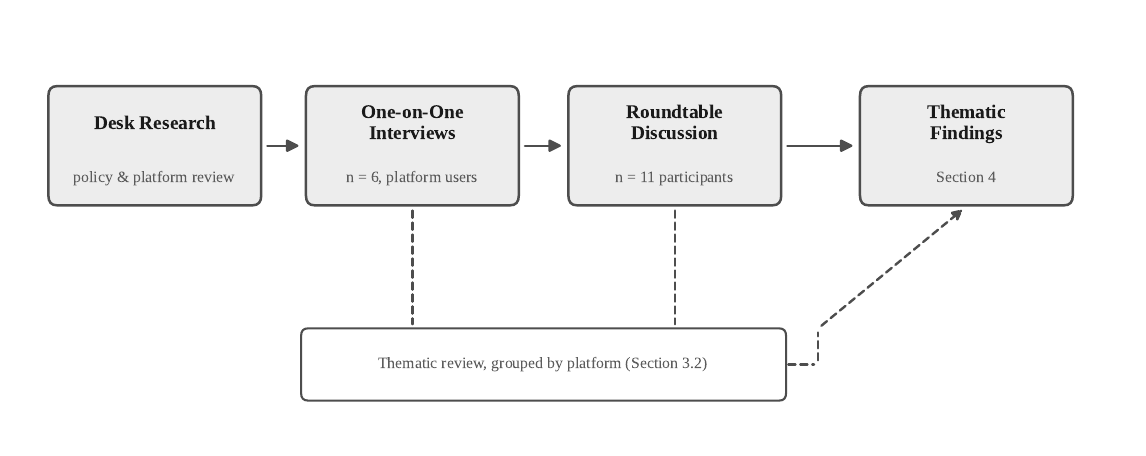}
  \caption{Research process, from desk research through one-on-one interviews and the roundtable discussion to the thematic findings reported in Section~\ref{sec:findings}.}
  \label{fig:methodology}
\end{figure}

This study uses a qualitative design. It combines one-on-one interviews with a facilitated roundtable discussion to synthesise first-hand accounts from users who encounter language and accessibility barriers directly, alongside feedback from user support staff on the platforms they support. Figure~\ref{fig:methodology} outlines this end-to-end research process, from the desk research that informed the interview and roundtable protocols, through the two data collection phases described in Section~\ref{subsec:sampling}, to the thematic findings reported in Section~\ref{sec:findings}.

\subsection{Participants and Sampling}
\label{subsec:sampling}
The eCitizen platform is intended for use by all members of the general public, so the potential participant pool was large. Respondent recruitment was based on convenience sampling, prioritising people who (a) had used at least one of the platforms under study, giving their accounts first-hand credibility, and (b) were reasonably accessible to the research team, either close to the interview location or able to attend the roundtable session which was held in-person.

Data collection took place in two phases. The first was six one-on-one interviews with users of the platforms most used by youth, since youth access was the study's central focus. The second was a roundtable with 11 participants, drawn from youth users aged 18 to 35 and user support staff from two neighbouring Huduma Centres. Both phases focused on four platforms considered essential to citizens' livelihoods: Ajira Digital, KRA iTax, HELB, and YEDF.

In total, 17 people took part, 2 user support staff and 15 users. Some users had experience of more than one platform, so the platform counts in Table~\ref{tab:participants} are not mutually exclusive. 3 participants had used Ajira Digital, 10 had used iTax, 12 had used HELB, and 1 had used YEDF.

\begin{table}[htbp]
  \centering
  \caption{Participants by role and platform of use.}
  \label{tab:participants}
  \begin{tabularx}{\textwidth}{X c c c c c}
    \toprule
    \textbf{Participant category} & \textbf{Ajira Digital} & \textbf{KRA iTax} & \textbf{HELB} & \textbf{YEDF} & \textbf{n} \\
    \midrule
    Platform users (overlapping; a user may have used more than one platform) & 3 & 10 & 12 & 1 & 15 \\
    User support staff (2 neighbouring Huduma Centres) & -- & -- & -- & -- & 2 \\
    \midrule
    \textbf{Total participants} & & & & & \textbf{17} \\
    \bottomrule
  \end{tabularx}
\end{table}

\subsection{Data Collection and Analysis}
Interviews and the roundtable covered participants' experience of language, literacy and accessibility barriers on their platform or platforms, and their views on what would make these platforms easier to use independently. Responses were reviewed thematically and grouped by platform to identify recurring barriers and suggestions, reported in Section~\ref{sec:findings}.

\subsection{Ethical Considerations}
Written consent was obtained from every participant before their interview or the roundtable session. All participants and their contributions are anonymised in this paper, using participant codes and role or platform descriptors rather than names. No named individual described in this paper is identified by their real name.

\section{Findings}
\label{sec:findings}
For Kenya's youth, accessibility problems on Ajira Digital, HELB and iTax translate to economic challenges. They limit the access citizens need to register businesses, apply for loans, and file taxes, and the barriers behind this run from language, through process design, to the assumptions built into the platforms themselves. The three themes these barriers can be classified into are language and literacy, process complexity, and structural barriers outside the interface.

\subsection{Language and Literacy}
Bureaucratic English-language prompts have no standard Swahili or any other native language equivalents, and on Ajira Digital, this compounds with content that is not personalised in the first place. \textit{P-Ajira1}, a community ambassador and youth user, described course text that is long, dense and poorly suited to new users or those with reading difficulties, and pointed, half in jest, to the `TikTok and YouTube Shorts era' as evidence that shortened attention spans call for multimodal content, images and short video, rather than text alone. He suggested automated translation, including evolving local slang such as \textit{Sheng}, would help more citizens engage with content in their own language. The platform's lack of text-to-speech or auditory support makes the independence question concrete. As a campus club trainer, \textit{P-Ajira1} recalled having to read out onboarding steps and complete forms on behalf of a visually impaired student, and in the absence of built-in support, users are left depending on scarce specialised trainers, which denies independence and risks alienation.

\subsection{Process Complexity}
Complexity pushes users toward intermediaries they cannot fully trust, and the cost of that shows up differently on each platform. \textit{P-HELB1}, a first-time applicant, was forced to spend time and money at a cyber caf\'e because the mobile HELB forms were too long and complex, an experience echoed by other participants. Long queues at cyber caf\'es during application periods lead to hurried service and data entry errors, sometimes costing students their funding or landing them the wrong scholarship amount, and reliance on third parties to complete forms also raises questions about data privacy. On iTax, \textit{P-iTax1}, a tech support officer, reported that over half of all traffic still comes from cyber caf\'es, where taxpayers rely on attendants to explain policy with a real risk of being misled. She cited wrongly filed NIL returns, in place of legitimate exemptions, as a direct and costly result of the lack of simple educational material. Across the platform, repetitive login loops add to this, particularly when applying for documents such as passports or certificates of good conduct.

\subsection{Structural Barriers}
Some of what excludes citizens sits outside the interface altogether. The platforms are designed as though every user owns a smartphone, which alienates those who do not. Participants and support staff proposed solutions that anticipate the framework developed later in this paper. \textit{P-HELB2}, a HELB support officer, and \textit{P-iTax1} both supported translation and summarisation technology to reduce clients' dependence on third parties and the errors and privacy risks that come with it. Others proposed a conversational `govbot' modelled on the existing KRA WhatsApp chatbot, zero-rated USSD access for citizens without smartphones, a dedicated accessibility tab for users with disabilities, and in-line plain-language tooltips for bureaucratic terms.

Service officers were themselves divided on the future of AI in their roles. Those already familiar with AI tools were positive, expecting it to free up time for complex cases and data-driven service delivery, and pointed to institutions like the Kenya School of Revenue Administration (KESRA) as natural venues for upskilling, while others cautioned that integration could meet resistance from staff less able to adapt. These discussions point to the fact that the barrier to eGovernment access is tied to economic opportunity. A human-centred AI strategy that prioritises independence, simplification, and multimodal and multilingual support looks, on this evidence, like a meaningful investment in the nation's youth livelihoods.

\section{Discussion}
\label{sec:discussion}
This section aligns the findings in Section~\ref{sec:findings} against Kenya's and the African Union's AI strategies, weighs AI's potential against its limits, and sets out a WCAG-based lens for diagnosing eCitizen's design, which the recommendations in Section~\ref{sec:recommendations} build on.

\subsection{African Union Continental and Kenya National AI Strategies}
The shift to GenAI and language technologies fits Africa's digital ambitions. The African Union Continental AI Strategy \cite{aucontinental2024} prioritises the adoption of AI in the public sector to deliver efficient services and improve people's livelihoods, in line with Agenda 2063 \cite{deghetto2016}. It frames AI as a tool for economic transformation, and it stresses governance, inclusion, and the use of African datasets to keep results relevant and reduce bias.

Kenya's approach mirrors this. The Kenya National AI Strategy (2025--2030) \cite{kenyastrategy2025} rests on pillars including AI Digital Infrastructure, Data Ecosystems, and Inclusion, Trust and Societal Impact. Better public service delivery through automation is a named priority. Developing localised AI models and Natural Language Processing (NLP) for Kenyan languages is a strategic action under the Research and Innovation pillar, which lines up with the text translation and summarisation this paper recommends for eCitizen. Both strategies tie the success of AI in service delivery to trust-based, ethical governance, and to closing the existing digital divide.

\subsection{AI is not the Silver Bullet}
AI-driven tools could support youth livelihoods, but this potential is largely untapped because of skill gaps and access barriers. Realising it needs investment in accessible AI infrastructure, wider digital literacy, and stronger AI skills, backed by policy that ensures ethical adoption, responsible data governance, and fair labour practices in the gig economy. Without these, it will be hard to create conditions where youth can actually use AI.

AI-powered language technologies could address Kenya's digital language divide, but practical adoption faces real local constraints. Infrastructure for large-scale data processing is limited, and training AI models is costly in a developing technological landscape. On the user side, there is uncertainty around trust and willingness to use AI-driven tools for tasks like tax filing or licence applications. To make AI in eGovernance work, Kenya needs affordable computing infrastructure, wider digital literacy, and clear data protection laws so citizen data is handled securely and transparently.

An inclusive digital future needs public service employees upskilled to manage AI tools, alongside systems designed from a human-centred perspective, intuitive, practical, and able to adapt to local language and changing digital needs. But these technologies do not solve youth unemployment on their own. They only make services easier to access. The root causes of unemployment need deliberate action: more investment in education, an enabling environment for business, equitable tax regulation, policy reform, youth representation in decision-making, and equal access to resources. Accessibility measures may carry an initial cost, but they are a worthwhile investment in a nation that values equity in the digital age.

\subsection{WCAG-Informed Design Lens}
Most of Kenya's youth already have access to smartphones and laptops, which they could use to engage with government services. Smartphone penetration in Kenya stood at 80.8\% \cite{ca2025}. This is an opportunity to reach most citizens through digital platforms, but only if the platforms are designed for them. Addressing only the surface-level user problem is not enough to meet a challenge this complex. A human-centred, systems-minded design \cite{both2018} treats bureaucratic language and organisational process as part of the problem alongside individual user needs. Websites providing critical public services, such as those for eGovernance, should be simple and self-explanatory by default.

The Web Content Accessibility Guidelines (WCAG) 2.1, developed by the World Wide Web Consortium (W3C), organise accessibility around four principles known as POUR. Perceivable and Understandable are about how content is presented, so users can access and understand it. Operable and Robust are about how functions and navigation work, so different users can actually use the platform. This paper uses these four principles as a design lens for the recommendations in Section~\ref{sec:recommendations}, not as a conducted evaluation. Table~\ref{tab:wcag_mapping} outlines, based on the accounts in Section~\ref{sec:findings}, where eCitizen currently falls short of each POUR principle. This is an illustrative reading of the evidence gathered, not a formal conformance audit.

\begin{table}[htbp]
  \centering
  \caption{Mapping WCAG 2.1 to eCitizen.}
  \label{tab:wcag_mapping}
  \begin{tabularx}{\textwidth}{l X X}
    \toprule
    \textbf{WCAG Principle} & \textbf{Definition (WCAG)} & \textbf{eCitizen Deficit} \\
    \midrule
    Perceivable & Information and interface components must be presentable to users in ways they can perceive, visual or auditory. & Text-only interface. No audio support at all. \\
    Operable & Interface components and navigation must be operable. & Dense, bureaucratic English. Complex multi-step forms with no plain-language alternative. Repetitive login loops when applying for documents. \\
    Understandable & Information and the operation of the interface must be understandable. & Multi-step navigation that often needs cyber caf\'e assistance to avoid errors. No tutorials, and unresponsive official support when processes change. \\
    Robust & Content must work reliably across a wide range of user agents and user groups. & No integrated translation or summarisation tools. Design assumes universal smartphone ownership, with no low-bandwidth or USSD alternative. \\
    \bottomrule
  \end{tabularx}
\end{table}

Read against this diagnosis, Table~\ref{tab:features_mapping} maps the features proposed in Section~\ref{sec:recommendations} onto the same four POUR principles.

\begin{table}[htbp]
  \centering
  \caption{Proposed features mapped to WCAG 2.1 POUR principles.}
  \label{tab:features_mapping}
  \begin{tabularx}{\textwidth}{l l X}
    \toprule
    \textbf{POUR Principle} & \textbf{Proposed Feature} & \textbf{How it addresses the principle} \\
    \midrule
    Perceivable & Text-to-Speech (TTS) & Gives access to users with visual impairments or reading difficulties, who cannot use a text-only interface. \\
    Operable & Speech-to-Text (STT) & Lets citizens speak requests or enter information, helping users who struggle with typing, spelling, or small-screen input. \\
    Understandable & Text Translation \& Summarisation & Turns bureaucratic content into accessible Swahili and other native languages, and condenses long forms and policies into plain-language summaries. \\
    Robust & Agentic AI Architecture & Integrates across government databases, so the platform works reliably across the devices and assistive technologies its users bring to it. \\
    \bottomrule
  \end{tabularx}
\end{table}

\begin{figure}[htbp]
  \centering
  \includegraphics[width=0.8\linewidth]{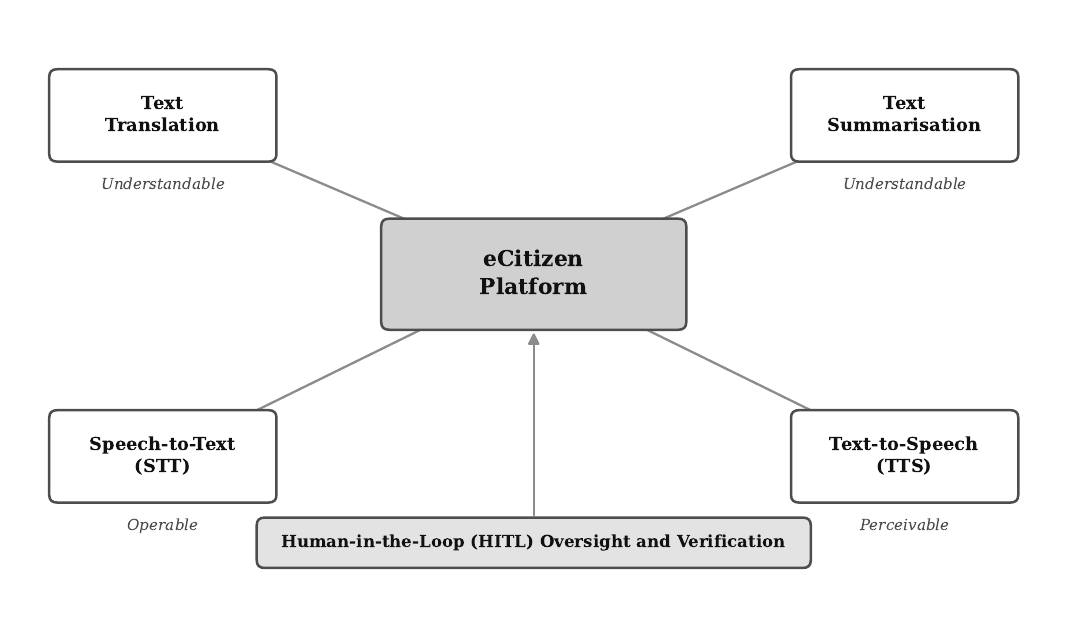}
  \caption{Proposed framework mapping the four AI-powered language technologies to the WCAG 2.1 (POUR) accessibility principles, embedded within eCitizen and subject to human-in-the-loop oversight.}
  \label{fig:framework}
\end{figure}

Figure~\ref{fig:framework} presents this same mapping between language technologies and the WCAG 2.1 POUR principles visually, situating the four features within eCitizen and under continuous human-in-the-loop oversight.

The adoption of these AI-powered language tools needs caution. GenAI systems in low-resource language environments, like many of Kenya's native languages, are prone to bias from unrepresentative training data, which can introduce inaccuracies and cultural insensitivity. These models can also hallucinate, generating false or misleading information, an unacceptable risk for services where accuracy matters to both the Perceivable and Understandable principles. Deploying these technologies safely needs continuous fine-tuning and human review during piloting and ongoing operation, so AI-generated outputs are checked and misinformation does not reach citizens using services such as tax filing, digital education, or business registration.

\section{Recommendations}
\label{sec:recommendations}
The findings in Section~\ref{sec:findings} and the WCAG diagnosis in Section~\ref{sec:discussion} point to five concrete actions. Together, they cover the data language technologies need, the content citizens actually read, the governance that keeps AI use accountable, the architecture that ties services together, and the people who support users day to day. Actionable recommendations for stakeholders are:

\begin{enumerate}
  \item \textbf{Fund foundational language data.} Public-private partnerships and academic institutions need funding to build high-quality, large-scale linguistic datasets for key local languages. The AI models discussed in this paper only work as well as the local data behind them, data that captures cultural and administrative nuance.
  \item \textbf{Mandate `Plain Language First'.} All new and updated digital public content should go through mandatory AI-driven text simplification before publication, with a target readability score set against cognitive accessibility standards such as PLAIN or E2R guidelines.
  \item \textbf{Establish an ethical AI governance mechanism.} A multi-stakeholder committee, including language experts and representatives from disability and low-literacy communities, should set ethical guardrails for AI use in translation and simplification, so legal precision and the original intent of the text are preserved.
  \item \textbf{Adopt an agentic AI architecture.} Procurement should move from simple, single-purpose chatbots to a comprehensive agentic AI platform that integrates across multiple government databases, letting citizens complete entire service transactions through one native-language, voice-enabled interface. The existing KRA WhatsApp chatbot is a useful local proof of concept for this kind of conversational interaction. Zero-rated USSD access should be prioritised alongside it, so citizens without a smartphone are not excluded from the same interface.
  \item \textbf{Upskill user support agents.} A technology-centred redesign needs user support staff trained alongside it, so users trust the system and can actually use the improved, more accessible platform.
\end{enumerate}

\begin{figure}[htbp]
  \centering
  \includegraphics[width=\linewidth]{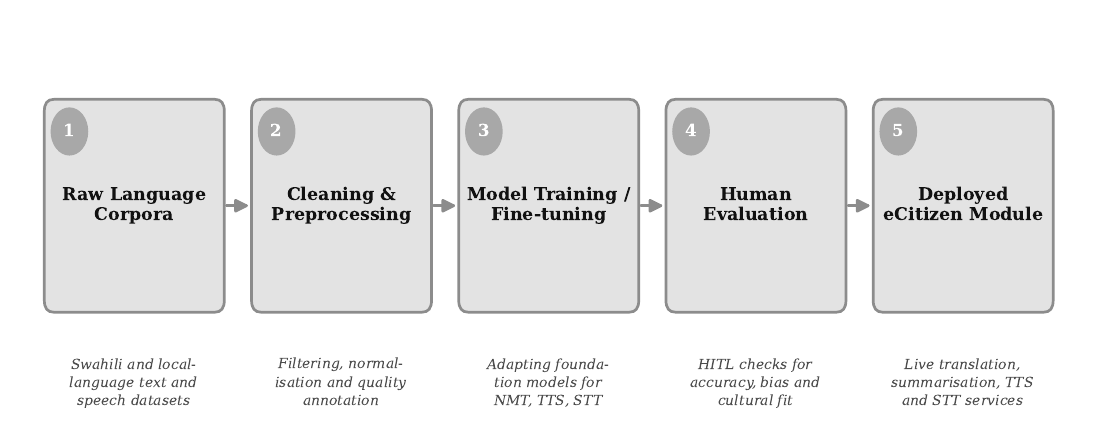}
  \caption{Simplified implementation pipeline, from raw language corpora to a deployed, usable eCitizen module.}
  \label{fig:pipeline}
\end{figure}

Figure~\ref{fig:pipeline} outlines in simplified form how the language technologies recommended above would move from raw corpora to a deployed, usable module within eCitizen: starting from Swahili and local-language datasets, through cleaning and fine-tuning of foundation models, to human evaluation and live deployment as translation, summarisation, TTS and STT services.

\section{Conclusion}
\label{sec:conclusion}
The eCitizen platform hosts over 22,000 services central to Kenyan citizens' livelihoods, but its English-only, text-only design excludes rural citizens, low-literacy users, and persons with disabilities. Interviews and a roundtable discussion with users and support staff of Ajira Digital, HELB, iTax and YEDF point broadly to three recurring barriers: language and literacy, process complexity that pushes users toward cyber caf\'es and intermediaries they cannot fully trust, and structural assumptions such as universal smartphone ownership. Read against WCAG 2.1, these barriers sit across all four POUR principles and point to interventions worth pursuing: automated translation and summarisation, text-to-speech and speech-to-text, and an agentic architecture, backed by funded language data, plain-language rules, ethical AI governance, and upskilled support staff. If deployed with human oversight, these interventions could improve access for citizens currently excluded from services they depend on.

\section{Limitations and Future Work}
\label{sec:limitations}
The findings in this paper are limited to the number of interviewees and the opinions they consented to share, since the seventeen participants accessible by the study are not a representative sample of eCitizen's user base. The WCAG principles are used here as a design lens rather than a conducted audit, and a structured WCAG 2.1 walkthrough of eCitizen and its affiliated platforms remains a priority for future work. Solutions proposed in Section~\ref{sec:recommendations} depend on the goodwill and coordinated action of government agencies, funders, and technology providers, none of which this paper can guarantee.

This paper also stops at the diagnostic and design stage: the framework in Figure~\ref{fig:framework} has not yet been implemented or tested against real eCitizen traffic. Future work will take this from framework to proof of concept in two stages. First, curating and releasing an open, anonymised version of the interview and roundtable data, together with the derived thematic codebook, so other researchers and platform teams can verify, extend or challenge these findings independently. Second, building and fine-tuning the NMT, text summarisation, TTS and STT models set out in Figure~\ref{fig:framework} on this released data, and integrating them into a working eCitizen prototype, to give this framework a technical proof of concept rather than a conceptual one, and to generate the accuracy, latency and human-in-the-loop review data needed before any production deployment.

\section*{Acknowledgments}
This study was conducted as part of \textit{AI and Livelihoods} research for The Pan-African Women PhD Scholars Collective, an initiative by Caribou and the Mastercard Foundation: \url{https://caribou.global/projects/the-pan-african-women-phd-scholars-collective-in-ai-livelihoods-2/}

Cynthia Amol wishes to thank Dr. Savita Bailur and Ms. Grace Natabaalo (Caribou), and the scholars at the collective, for their input. She also thanks the focus group respondents (names anonymised) for sharing their valuable experiences.

\bibliographystyle{unsrturl}
\bibliography{refrences}

\end{document}